\documentclass[%
 preprint,
 amsmath,amssymb,
 aps,prl,
]{revtex4-1}

\usepackage{graphicx,epstopdf,amssymb,amsfonts,amsmath,amsthm,array,
mathrsfs,amscd}
\usepackage{dcolumn}
\usepackage{bm}

\def\be{\begin{equation}}\def\ee{\end{equation}}

\def\cvp{\raise 2pt\hbox{,}} 
 \def\tr{\mathop{\rm tr}\nolimits}

 \def\d{{\rm d}}

\def\plb#1#2#3{{\it Phys.\ Lett.\ }{\bf B #1} (#2) #3}
\def\npb#1#2#3{{\it Nucl.\ Phys.\ }{\bf B #1} (#2) #3}

\def\prl#1#2#3{{\it Phys.\ Rev.\ Lett.\ }{\bf #1} (#2) #3}

\def\cmp#1#2#3{{\it Comm.\ Math.\ Phys.\ }{\bf #1} (#2) #3}

\def\mpla#1#2#3{{\it Mod.\ Phys.\ Lett.\ }{\bf A #1} (#2) #3}

\def\imath#1#2#3{{\it Invent math }{\bf #1} (#2) #3}
\def\jdiffgeo#1#2#3{{\it J.\ Diff.\ Geom.\ }{\bf #1} (#2) #3}

\begin{document}


\title{Random Geometry, Quantum Gravity and the K\"ahler Potential
}

\author{Frank Ferrari and Semyon Klevtsov}
 \altaffiliation[Also at ]{ITEP, Moscow, Russia.}
 \email{frank.ferrari@ulb.ac.be, semyon.klevtsov@ulb.ac.be}
 \affiliation{Service de 
 Physique Th\'eorique et
 Math\'ematique, Universit\'e Libre de Bruxelles and International
 Solvay Institutes, Campus de la Plaine, CP 231, B-1050 Bruxelles,
 Belgium}
\author{Steve Zelditch}%
 \email{zelditch@math.northwestern.edu}
\affiliation{
 Northwestern University, Evanston, IL 60208, USA
}%

\date{\today}

\begin{abstract}

We propose a new method to define theories of random geometries, using
an explicit and simple map between metrics and large hermitian
matrices. We outline some of the many possible applications of the
formalism. For example, a background-independent measure on the space
of metrics can be easily constructed from first principles. Our framework suggests the relevance of a new gravitational effective action and we show that it occurs when coupling the massive scalar field to two-dimensional gravity. This yields new types of quantum gravity models generalizing the standard Liouville case.

\end{abstract}

\maketitle


\section{\label{s1} Introduction}

Defining and studying theories of random surfaces \cite{pol1}, or more generally random geometries, is a fundamental problem in science with a wide
range of applications, from string theory and quantum gravity to
statistical physics and probability theory. Most of the literature has
focused on random surfaces. In this case, the problem simplifies because 
any metric $g$ can be put into a simple form using diffeomorphisms, the so-called conformal gauge,
\begin{equation}
\label{confgauge}
g=e^{2\sigma}g_{0}\, ,
\end{equation}
where $g_{0}$ is some reference background metric which depends on at most a finite number of parameters. The standard route to define random metrics is then to consider that the unconstrained scalar $\sigma$ is a gaussian free field \cite{DDK}. One obtains in this way a family of non-trivial models, the so-called Liouville theories, that are parametrized by a single positive constant multiplying the Liouville action
\be\label{Liouvilleaction} S_{\text{L}}(g_{0},g) = \int\!\d^{2}x\,\sqrt{g_{0}}\bigl( g_{0}^{ab}\partial_{a}\sigma\partial_{b}\sigma + R_{0}\sigma\bigr)\, .\ee
These models are natural two-dimensional generalizations of the brownian random paths. They are known to be related to the continuum limit of discretized versions of random geometries formulated using double-scaled matrix models \cite{kaza}. They are also of interest for probabilists who have recently proved a rigorous version 
\cite{sheff} of the KPZ relation \cite{KPZ}.

In spite of its many successes, the Liouville formulation suffers from a certain number of shortcomings. For example, one of the fundamental application of the model has been to the theory of two-dimensional quantum gravity. A crucial requirement in quantum gravity is background-independence which, in the present context, is equivalent to independence with respect to the reference metric $g_{0}$ in \eqref{confgauge}. However, the Liouville theory is background-dependent, because both the Liouville action \eqref{Liouvilleaction} and the path integral measure on the Liouville field, which is derived from the following metric in field space,
\be
\label{Liouvillemetric}
\lVert\delta\sigma\rVert_{0}^{2} =\int_{\Sigma}\!\d^{2}x\,\sqrt{g_{0}}\,(\delta\sigma)^{2}\, ,\ee
are background-dependent. A major result in Liouville theory \cite{DDK} is to demonstrate that the model can actually be made background-independent when coupled to a conformal field theory of central charge $c\leq 1$ by adjusting the parameter multiplying the action \eqref{Liouvilleaction}. The cases $c>1$ remain open, but even when $c\leq 1$ it would be desirable to understand why the Liouville model is the correct theory of quantum gravity. From first principle, the background-independent path integral measure $\mathscr D\sigma$ should be derived from the background-independent metric on the space $\mathscr M$ of all two-dimensional metrics
\be
\label{Calabimetric}
\lVert\delta\sigma\rVert^{2} = \int_{\Sigma}\!\d^{2}x\,\sqrt{g}\,(\delta\sigma)^{2}=\int_{\Sigma}\!\d^{2}x\,\sqrt{g_{0}}\,e^{2\sigma}(\delta\sigma)^{2}\, .\ee
This formula is very unusual, because it depends non-linearly on the field $\sigma$, and it is not clear a priori how to use it to build a path integral.

In this letter, we propose a new approach to the theory of random geometry, based on the profound geometrical properties of the space of metrics on a K\"ahler manifold \cite{Mabuchi, Mabuchi2, Tian}. In this framework, we can construct very simple regularized versions of 
\eqref{Calabimetric} and $\mathscr D\sigma$. A wealth of new theories of random metrics can be considered. In particular, it is natural to study models for which the gravitational action is given by the Mabuchi functional 
\cite{Mabuchi}, which is singled out by its unique geometrical properties. Remarkably, we show that the effective gravitational action for a massive scalar field coupled to gravity in two dimensions does contain the Mabuchi action on top of the standard Liouville term, yielding new quantum gravity models with profound geometrical features.

We shall restrict ourselves in the following to the simplest case of random metrics on the two dimensional sphere $\Sigma=\text{S}^{2}$. We can then choose the background metric in \eqref{confgauge} to be the round metric of area $A_{0}$,
\be\label{roundmet} g_{0}=\frac{A_{0}}{\pi}\frac{|\d z|^{2}}{(1+|z|^{2})^{2}}\,\cvp\ee
where $(z,\bar z)$ are the standard stereographic coordinates. Generalizations to arbitrary Riemann surfaces $\Sigma$ and to higher dimensional K\"ahler manifolds are presented elsewhere \cite{FKZ1,FKZ2}. 

\section{\label{s2} The K\"ahler potential and the construction of
metrics}

Our starting point is to write the conformal factor $e^{2\sigma}$ in terms of the K\"ahler potential $\phi$ defined by the equation
\be\label{phidef} e^{2\sigma} = \frac{A}{A_{0}} -\frac{1}{2}A\,\Delta_{0}\phi\, ,
\ee
where $A$ is the area for the metric $g=e^{2\sigma}g_{0}$ and $\Delta_{0}$ the positive laplacian for the metric $g_{0}$. Equation \eqref{phidef} can always be solved for $A$ and $\phi$ in terms of $\sigma$, and the solution is unique up to constant shifts in $\phi$. \emph{We propose to focus on $\phi$ instead of $\sigma$ to define random metrics.}

The field $\phi$ must satisfy the fundamental inequality
\be\label{constphi} \Delta_{0}\phi < 2/A_{0}\ee
coming from the positivity of the metric $g$. To define a path integral over $\phi$, we must regularize the theory by introducing a UV cut-off and also solve the constraint \eqref{constphi}. Remarkably, this can be done in a very elegant way.

A simple method to regularize is to expand the field on spherical harmonics up to spin $N$, 
corresponding to a short distance cut-off $\ell\sim A^{1/2}/N$, and then take the limit $N\rightarrow\infty$. In stereographic coordinates, it is not difficult to see that a basis for the space of all spherical harmonics of spin up to $N$ is given by the functions $f_{ij}= \bar s_{i}(\bar z) s_{j}(z) \lambda_{0}^{2}(z,\bar z)$, $0\leq i,j\leq N$, where the $(s_{i})_{0\leq i\leq N}$ forms a basis for the holomorphic polynomials of degrees up to $N$ and $1/\lambda_{0}^{2} = (1+|z|^{2})^{N}$. It is actually convenient to choose
\be\label{sidef} s_{j}(z) = \sqrt{\frac{N!}{j!(N-j)!}}\, z^{j}\, ,\quad 0\leq j\leq N\, .\ee
In order to make the constraint \eqref{constphi} tractable, the idea, which is based on profound methods in K\"ahler geometry \cite{Tian}, is to expand
\be\label{phiexp} e^{2\pi N\phi_{N}} = \sum_{0\leq i,j\leq N}\lambda_{0}^{2}\,\bar s_{i}(\bar z) H_{ij} s_{j}(z)\ee
instead of $\phi$ itself. The field $\phi_{N}$ is the regularized version of $\phi$ and the associated metric is called a Bergman metric. The matrix $H$ in \eqref{phiexp} must be hermitian since the K\"ahler potential is real. It is defined up to multiplication by strictly positive constants, since constant shifts in the K\"ahler potential are immaterial. We can thus impose the condition $\det H=1$. Moreover, \emph{$H$ must be positive-definite.} This ensures that the right hand side of eq.\ \eqref{phiexp} is strictly positive, and, less trivially, also ensures that \eqref{constphi} is automatically satisfied, as a little calculation using the Cauchy-Schwarz inequality shows. The converse statement is also true \cite{Tian}: any $\phi$ satisfying \eqref{constphi} can be approximated by an expansion like \eqref{phiexp}, for $N$ large enough, where $H$ is a positive-definite hermitian matrix given by
\be\label{Hilbmap} H^{-1}_{ji} = \int_{\Sigma}\!\d^{2}x\,\sqrt{g}\, \lambda_{0}^{2}e^{-2\pi N\phi}\bar s_{i} s_{j}\, .\ee
The fact that $\phi_{N}$, given by \eqref{phiexp}, converges to $\phi$ when $N\rightarrow\infty$ if $H$ is given by \eqref{Hilbmap} (with the same convergence property being true for the associated metric and all its derivatives) can be interpreted in terms of natural properties of the lowest Landau level for a charged particle on $\text{S}^{2}$ in a magnetic field of strength $\sim N\rightarrow\infty$. We cannot expand on this interesting point here but more details can be found in \cite{DK, FKZ2}.

The important conclusion is that the symmetric space $\mathscr M_{N}=\text{SL}(N+1,\mathbb C)/\text{SU}(N+1)$ of $(N+1)\times (N+1)$ positive-definite hermitian matrices $H$ of determinant one provides a regularization of the space $\mathscr M$ of all metrics on $\text{S}^{2}$. Metrics in $\mathscr M_{N}$ are parametrized by $H$ through the formulas \eqref{confgauge}, \eqref{phidef}, \eqref{phiexp} and the space $\mathscr M_{N}\times\mathbb R_{+}^{*}$, where the $\mathbb R_{+}^{*}$ factor parametrizes the total area, goes to the space of all metrics $\mathscr M$ when $N\rightarrow\infty$. A theory of random metrics can then be defined by choosing some probability measure on the space $\mathscr M_{N}$, corresponding to a particular matrix model. The continuum limit is associated with the large $N$ limit. Let us emphasize that the integral over angles is crucial in these models, because the relevant observables depend on the full matrix $H$ and not only on its eigenvalues, as, for example, the formulas \eqref{phiexp} and \eqref{Hilbmap} clearly show.

\section{\label{s3} Background-independent measure}

The metric \eqref{Calabimetric} expressed in terms of the variables $A$ and $\phi$ defined in \eqref{phidef} reads
\be\label{Calabimet2}
\lVert\delta\sigma\rVert^{2} = \frac{(\delta A)^{2}}{4A} + \frac{A^{2}}{16}\int\!\d^{2}x\,\sqrt{g}\, (\Delta_{g}\delta\phi)^{2}\, ,\ee
where $\Delta_{g}$ is the laplacian for the metric $g$. If we introduce the natural metric on the space of K\"ahler potentials $\phi$
\be\label{Mabushimet}\lVert\delta\phi\rVert^{2} = 
\int\!\d^{2}x\,\sqrt{g}\, (\delta\phi)^{2}\, ,\ee
then the background-independent measures $\mathscr D\phi$ and $\mathscr D\sigma$ associated to \eqref{Calabimetric} and \eqref{Mabushimet} respectively are simply related,
\be\label{measurerel} \mathscr D\sigma = A^{1/6}\d A\, \mathscr D\phi\, {\det}'(A\Delta_{g})\, ,\ee
and thus the problems of defining $\mathscr D\sigma$ and $\mathscr D\phi$ are equivalent. The determinant ${\det}'(A\Delta_{g})$ of the laplacian of the metric of unit area $g/A$ excludes the constant zero mode because the constant shifts in $\phi$ are unphysical. It is defined as usual via the zeta-function regularization procedure. The overall power of $A$ in \eqref{measurerel} is derived in general from a one-loop calculation and the result indicated in \eqref{measurerel} corresponds to the Liouville theory on the sphere \cite{FKZ2}.

In the context of the regularized theory defined in the previous Section, the choice of background is related to the choice of a metric associated to the identity matrix $H=\mathbb{I}$, or equivalently to a choice of basis $(s_{j})_{0\leq j\leq N}$. For example, with the choice \eqref{sidef}, $H=\mathbb{I}$ corresponds to the round metric \eqref{roundmet}.
Changing the background amounts to changing $H$ into $MHM^{\dagger}$ for some invertible matrix $M$. The metric on $\mathscr M_{N}$ defined by
\be\label{metMN} \lVert\delta H\rVert_{N}^{2} = \tr (H^{-1}\delta H)^{2}\, ,\ee
as well as its associated volume form, $\mathscr D_{N}H$ are thus manifestly background-independent.

Clearly, since at large $N$ the space $\mathscr M_{N}\times\mathbb R_{+}^{*}$ approximates the space $\mathscr M$ of all metrics, the measure $\mathscr D_{N}H$ must yield a regularized version of the background-independent measure on $\mathscr M$. This can be explicitly checked by studying the large $N$ asymptotics of \eqref{metMN}, again using the properties of the associated lowest Laudau level problem. The result
\be\label{dphidH} \lVert\delta\phi\rVert^{2}/A = \lim_{N\rightarrow\infty} \lVert\delta H\rVert_{N}^{2}/(8\pi^{2}N^{3})\ee
shows that $\mathscr D_{N}H$ provides a definition of $\mathscr D\phi$ at large $N$. Path integrals over metrics are then defined by
\be\label{defpath} \int_{\mathscr M}\!\mathscr D\phi\, e^{-S(\phi)}\sim\lim_{N\rightarrow\infty}\int_{\mathscr M_{N}}\!\mathscr D_{N}H\, e^{-S(\phi_{N}(H))}\, ,\ee
where $\phi_{N}(H)$ is given by \eqref{phiexp} and $\sim$ means that suitable rescalings (renormalizations) need to be performed when taking the limit.


%
\section{\label{s4} A new effective action for gravity}

The Liouville action first entered the field of two-dimensional quantum gravity because it provides an integrated version of the conformal anomaly \cite{pol1}. This implies that the full metric dependence of the partition function of a matter \emph{conformal field theory} of central charge $c$ coupled to gravity can be easily found. Taking into account the ghost CFT coming from the gauge fixing \eqref{confgauge}, the path integral over metrics reduces to
\be\label{qg1} \int_{\mathscr M}\!\mathscr D\sigma \, e^{\frac{c-26}{24\pi} S_{\text L}(g_{0},g)} Z(g_{0})\, ,\ee
where $Z(g_{0})$ is the partition function of the matter plus ghost system. The model thus reduces to the study of two decoupled CFT, the matter/ghost theory and the Liouville theory. Background independence is equivalent to the fact that the total central charge must be zero.

What happens when one couples a matter field theory which is \emph{not} conformal to gravity has been much less studied, in particular in the continuum formalism (see e.g.\ \cite{Ising} for the Ising model on the lattice, and \cite{Zam} for a perturbative approach in the continuum formalism). The obvious difficulty is that the metric dependence of the matter partition function can no longer be deduced from the conformal anomaly, and thus the effective gravitational action $S_{\text{eff}}$, defined by
\be\label{graveff} Z(g) = e^{-S_{\text{eff}}(g_{0},g)}Z(g_{0})\, ,\ee
is no longer given by the Liouville action. In particular, we expect 
$S_{\text{eff}}$ to be non-local in the Liouville field $\sigma$. Nevertheless, some fundamental properties, already present in the CFT case, must remain valid. First, background independence implies that the total gravity plus ghost plus non-conformal matter system must still be a conformal field theory of vanishing total central charge. Second, \eqref{graveff} implies that $S_{\text{eff}}$ must satisfy the following one-cocycle consistency conditions,
\begin{align}\label{cocycle1} S_{\text{eff}}(g_{1},g_{2}) & = -S_{\text{eff}}(g_{2},g_{1})\, ,\\
\label{cocycle2} S_{\text{eff}}(g_{1},g_{3}) &= S_{\text{eff}}(g_{1},g_{2}) + S_{\text{eff}}(g_{2},g_{3})\, .\end{align}

The above conditions are non-trivial but, unlike the CFT case, we cannot expect to derive from them a universal formula for $S_{\text{eff}}$. Let us thus 
simplify the problem by studying an expansion when the mass scale governing the non-conformality of the matter theory is small. Then we may expect to see some universality emerging, at least at leading order. For example, we consider a massive scalar field $X$ with action
\be\label{massivescalar} S_{\text{m}} = \frac{1}{8\pi} \int\!\d^{2}x\, 
\sqrt{g}\,\bigl( g^{ab}\partial_{a}X\partial_{b}X + q R X+ m^{2}X^{2}\bigr)\, ,\ee
where as usual $R$ is the Ricci scalar and $q$ an arbitrary dimensionless parameter, in the small $m^{2}$ expansion. This expansion is non-perturbative, because the mass term $m^{2}X^{2}$ is not a well-defined operator in the CFT at $m=0$. At fixed area, it will be valid for $m^{2}A\ll 1$. If we want to integrate over areas, then the cosmological constant should be chosen to be much larger than $m^{2}$. At leading order, as is usual in a small mass expansion, we may expect the effective action to be non-local with respect to $\sigma$, with terms of the form $A^{-1}\partial^{-2}\sigma$ or $m^{2}\partial^{-2}\sigma$ which, from \eqref{phidef}, may be made local in terms of $\phi$.

It turns out that there does exist an extremely natural functional of $\phi$, the so-called Mabuchi action \cite{Mabuchi}. It is given on the sphere by
\begin{equation}\label{Mabuchidef} S_{\text M}(g_{0},g) = \int_{\text{S}^{2}}\!\d^{2}x\, \sqrt{g_{0}}\bigl(-2\pi g_{0}^{ab}\partial_{a}\phi\partial_{b}\phi + 8\pi\phi/A_{0}- R_{0}\phi + 4\sigma e^{2\sigma}/A\bigr)\, ,\end{equation}
%
%
%
where the metrics $g$ and $g_{0}$ are related by \eqref{confgauge} and \eqref{phidef}.
The Mabuchi action satisfies all the required consistency conditions and actually shares many properties with the Liouville action. It is invariant under constant shifts of $\phi$, and thus well-defined on the space of metrics. It satisfies the cocycle conditions \eqref{cocycle1} and \eqref{cocycle2}, as can be checked straightforwardly. It is bounded from below and is convex in the metric \eqref{Mabushimet}, making it a suitable candidate to be used as an action in a path integral. It has a unique minimum, corresponding to the metric of constant scalar curvature, which is the round metric \eqref{roundmet} in the case of the sphere. It admits higher dimensional generalizations with similar properties. For these 
reasons it plays a central r\^ole in geometry, in particular in the study of constant scalar curvature metrics on K\"ahler manifolds. Finally, a regularized version of $S_{\text M}$ can be naturally constructed \cite{dondet,FKZ1}.

We have computed explicitly the effective gravitational action for the massive scalar field \eqref{massivescalar}, to leading order in the small 
mass expansion. It is convenient to add into the model a spectator CFT of central charge $c$. On the sphere and at fixed area, and when $q\not =0$, the result, up to terms of order $m^{2}A$, then reads
%
%
%
\be\label{Seffscalar} S_{\text{eff}}(g_{0},g) = \frac{25-3q^{2}-c}{24\pi}
S_{\text L}(g_{0},g) + \frac{q^{2}}{4} S_{\text M}(g_{0},g)\, .\ee
When $q=0$, the leading correction to the Liouville term, given by $\frac{m^{2}A}{16\pi} S_{\text M}(g_{0},g)$, is also governed by the Mabuchi action. It is also possible to consider correlation functions, for example
\be\label{corrfunc} \Bigl\langle\prod_{j}\int\!\d^{2}x_{j}\,\sqrt{g}\, e^{ik_{j}X(x_{j})}\Bigr\rangle\, .\ee
The gravitational dressing then involves new factors depending on the Mabuchi action as well as on a closely related functional called the Aubin-Yau action. Full details on these results are presented in \cite{FKZ2}.

A startling possibility is to adjust the parameters such that the Liouville contribution to the gravitational action vanishes altogether. Since the most natural measure in the present context is $\mathscr D\phi$ instead of $\mathscr D\sigma$, and the determinant factor in \eqref{measurerel} is equivalent to the contribution of a $c=-2$ CFT, this is achieved when $27-c-3q^{2}=0$ in our model. We are then left with an entirely new quantum gravity path integral, 
\be\label{MabQG} \int_{\mathscr M}\!\mathscr D\phi\, e^{-q^{2}S_{\text M}/4}\, ,\ee
which can be regularized and studied using the tools presented in the previous sections, see e.g.\ \eqref{defpath}. Note that the coupling $q^{2}$ in front of the Mabuchi action can be made arbitrary and thus the strength of the quantum fluctuations are chosen at will.

\section{Conclusion}

The approach to the theory of random metrics that we have proposed provides a new point of view with a deep interplay between matrix models and geometrical techniques. It allows to address fundamental questions in two dimensional quantum gravity and suggests new interesting models. It also opens a window on the theory of higher dimensional fluctuating geometries. We hope it will provide many fruitful insights into these hard but fundamental issues in theoretical physics.

\section{Acknowledgments}

We would like to thank M.~Douglas for useful discussions.

This work is supported in part by the belgian FRFC (grant 2.4655.07), the belgian IISN (grant 4.4511.06 and 4.4514.08), the IAP Programme (Belgian Science Policy), the RFFI grant 11-01-00962 and the NSF grant DMS-0904252.

\end{document}